\documentclass{IEEEtran}
\usepackage{cite}
\usepackage{amsmath,amssymb,amsfonts}
\usepackage{graphicx}
\usepackage{textcomp,nicefrac}
\usepackage{booktabs}
\usepackage{subcaption}
\usepackage{url}

\def\BibTeX{{\rm B\kern-.05em{\sc i\kern-.025em b}\kern-.08em
T\kern-.1667em\lower.7ex\hbox{E}\kern-.125emX}}

\begin{document}

\title{A Congestion-Aware RDMA Front-End for LST Advanced Camera Readout}

\author{F.~Marini, M.~Bellato, A.~Bergnoli, A.~Griggio, R.~Isocrate, L.~Modenese, F.~Montecassiano and M.~Toffano%
\thanks{This work has been submitted to the IEEE for possible publication. Copyright may be transferred without notice, after which this version may no longer be accessible.}%
\thanks{F. Marini, M. Bellato, A. Bergnoli, A. Griggio, R. Isocrate, L. Modenese, F. Montecassiano are with INFN Section of Padova, Padua, Italy (e-mail: filippo.marini@pd.infn.it).}
\thanks{M. Toffano is with INFN Section of Padova, Padua, Italy and with University of Padova, Padua, Italy}}

\maketitle

\begin{abstract}
The Advanced Camera (AdvCam) upgrade of the Large-Sized Telescopes (LSTs) calls for a fully digital readout in which each front-end board streams data to the event-building servers over commercial 10\,Gb/s Ethernet, using an FPGA implementation of Remote Direct Memory Access (RDMA) over Converged Ethernet v2 (RoCEv2). In a previous work, a customized RoCEv2 core integrated with the SLAC SURF UDP/MAC stack was shown to sustain near-line-rate data transfers from a Xilinx KCU105 to a server equipped with a Mellanox ConnectX-5 NIC. However, the AdvCam topology is an incast funnel: many 10\,GbE links converge onto a single uplink, and bare Priority Flow Control (PFC) congestion management protocol is known to be unfair and to suffer from head-of-line blocking. This work extends the previous design with the Data Center Quantized Congestion Notification (DCQCN) algorithm. A protocol-level ns-3 simulation is used to validate the DCQCN dynamics and to explore the parameter space; an FPGA rate-limiter integrated into the SURF MAC reacts to incoming Congestion Notification Packets (CNPs) according to the DCQCN state machine; a cocotb testbench with an emulated Python switch verifies the firmware end-to-end against a Soft-RoCE receiver; and a three-sender incast hardware test on a DCQCN-enabled commercial switch reproduces the canonical fair-share, reactive-redistribution, and saw-tooth behaviour expected from a commercial RDMA NIC. The internal rate-limiter register is shown to track the measured RoCEv2 throughput cycle by cycle.
\end{abstract}


\section{Introduction}
\label{sec:intro}
\IEEEPARstart{T}{he} Large-Sized Telescopes (LSTs) are designed to capture the lowest-energy gamma rays observed by ground-based imaging atmospheric Cherenkov telescope arrays. Their Advanced Camera (AdvCam) upgrade~\cite{intro2}, based on Silicon Photo Multipliers (SiPMs), is expected to increase the number of channels by a factor of four (to about 8000 pixels per camera) and to raise the data throughput by roughly an order of magnitude relative to the legacy PhotoMultiplier Tube (PMT) design~\cite{intro}. To cope with this throughput while keeping the on-detector electronics affordable and serviceable, the AdvCam readout adopts a fully digital architecture in which the front-end board itself performs digitization, local triggering, packetization, and direct RDMA WRITE transfer to the event-building servers over a commercial Ethernet fabric, with no intermediate custom backend layer.

\begin{figure}[t]
\centering
\includegraphics[width=0.95\columnwidth]{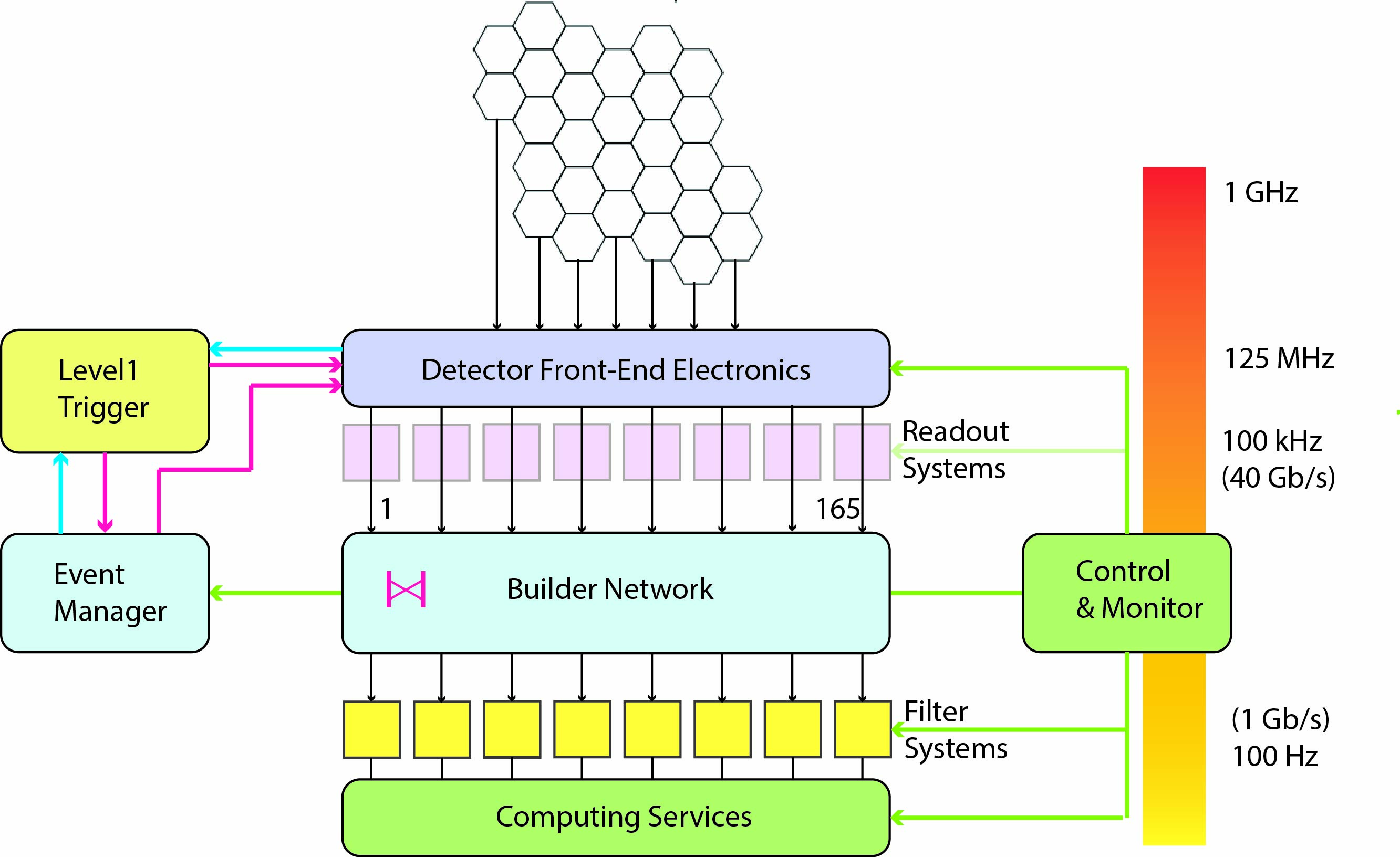}
\caption{Typical HEP DAQ architecture. In the AdvCam design the dedicated readout-system stage (purple boxes) is bypassed: front-end data flow directly from the front-end electronics to the builder network.}
\label{fig:daq_topology}
\end{figure}

Fig.~\ref{fig:daq_topology} sketches a conventional HEP-like DAQ topology that the AdvCam architecture will follow. The novelty is that the dedicated \emph{readout system} --- conventionally a layer of custom backend cards with many high-speed input links and few aggregated output links~\cite{cms} --- is bypassed: the front-end boards themselves perform the digitization, triggering, and packetization that the readout backends would otherwise carry out, and push the resulting data directly to the DAQ servers via RoCEv2 over a commercial Ethernet fabric.

The hardware demonstrator of this architecture and a customized FPGA implementation of RoCEv2~\cite{ibta2014rocev2}, written in Bluespec SystemVerilog~\cite{bsv} and integrated with the SLAC SURF UDP/MAC stack~\cite{surf}, have been described in detail in~\cite{marini2025fpga}. There, the firmware was shown to sustain $\sim$9.7\,Gb/s, essentially at the line rate of a 10\,Gb/s link, in a point-to-point RDMA WRITE test against a server equipped with a Mellanox ConnectX-5 NIC~\cite{connectx5}. That work, however, deliberately left open one essential question for a realistic deployment: in a many-to-one (incast) configuration like the AdvCam DAQ funnel, how is congestion handled?

RDMA assumes a lossless fabric: a single dropped packet triggers a Go-Back-N retransmission~\cite{ibta2014rocev2} that collapses throughput. The classical mechanism used to keep RoCEv2 fabrics lossless is Priority Flow Control (PFC)~\cite{ieee8021qbb}, but PFC is known to be unfair at the per-flow level and to suffer from head-of-line blocking~\cite{guo2016rdma}. The Data Center Quantized Congestion Notification (DCQCN) algorithm~\cite{zhu2015dcqcn} was introduced to address these limitations. It is now the standard congestion-control algorithm in commercial RDMA deployments and is supported, with varying levels of compliance, by all major vendors of RDMA-capable NICs.

This paper describes the addition of DCQCN to the FPGA RoCEv2 stack of~\cite{marini2025fpga}. The contributions are: (i) a protocol-level ns-3~\cite{ns3} validation of the chosen DCQCN parameter set on an incast topology representative of the AdvCam fabric; (ii) an FPGA rate-limiter module integrated with the SURF MAC, with run-time-programmable DCQCN parameters; (iii) an end-to-end firmware simulation flow based on cocotb~\cite{cocotb} and a Python emulated switch; and (iv) hardware measurements on a three-sender incast test bed built around a DCQCN-enabled commercial switch and a ConnectX-5 receiver, reproducing the qualitative and quantitative dynamics expected from DCQCN.

\section{The DCQCN Algorithm}
\label{sec:dcqcn}
DCQCN is an end-to-end, per-flow, rate-based congestion-control algorithm originally proposed for large-scale RDMA deployments. The algorithm is described exhaustively in~\cite{zhu2015dcqcn}; the elements relevant to the FPGA implementation are recalled in the following.

\begin{figure}[t]
\centering
\includegraphics[width=0.75\columnwidth]{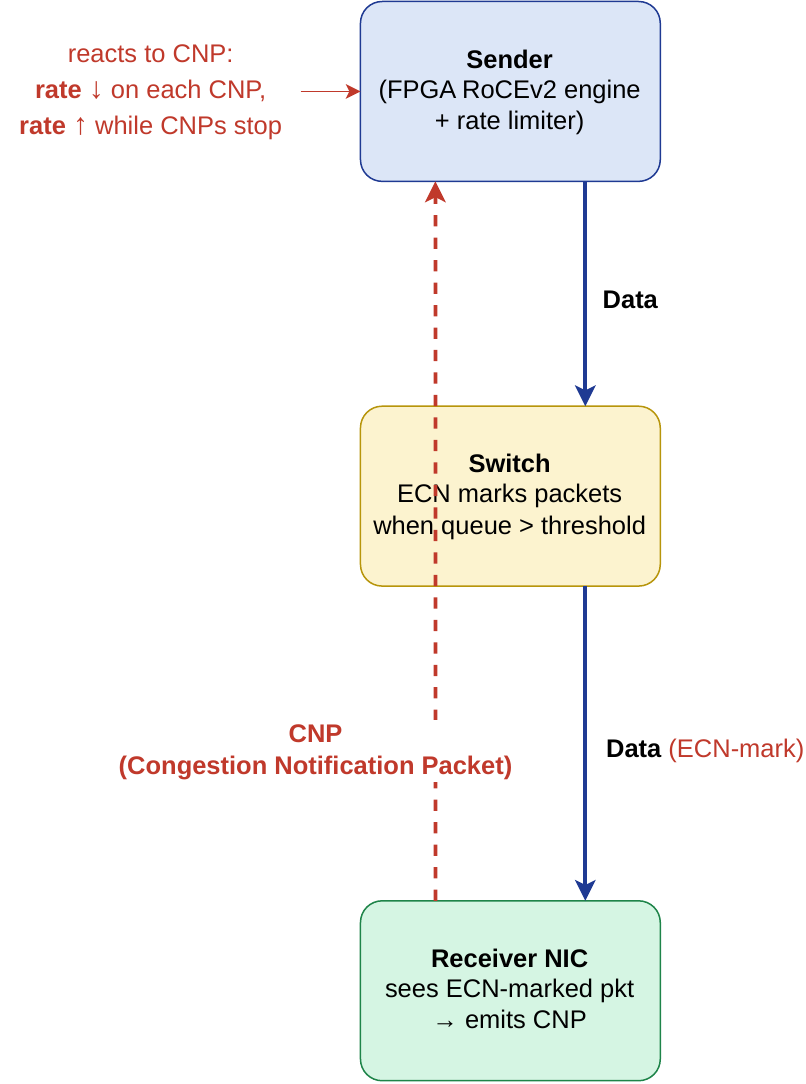}
\caption{DCQCN feedback loop: the switch ECN-marks, the receiver replies with a CNP, the sender reacts.}
\label{fig:dcqcn_loop}
\end{figure}

The DCQCN feedback loop, shown schematically in Fig.~\ref{fig:dcqcn_loop}, involves three actors. The \emph{switch} marks data packets with the Explicit Congestion Notification (ECN) flag when its egress queue exceeds a configurable threshold, without dropping them and without raising PFC. The \emph{receiver} (Notification Point), upon seeing an ECN-marked packet, replies with a Congestion Notification Packet (CNP) to the offending sender. The \emph{sender} (Reaction Point) decreases its rate sharply on each CNP and recovers gradually while no further CNPs arrive.

Quantitatively, the sender maintains a congestion estimator $\alpha \in [0,1]$, a current rate $R_C$, which acts as the ceiling of the front-end's egress throughput, and a target rate $R_T$. On each received CNP, the estimator is updated as
\begin{equation}
\alpha \leftarrow (1-g)\,\alpha + g, \qquad 0 < g \ll 1
\label{eq:alpha_up}
\end{equation}
where $g$ is the \textit{gain} parameter set by the user, while in CNP-free intervals it decays toward zero with the same gain:
\begin{equation}
\alpha \leftarrow (1-g)\,\alpha.
\label{eq:alpha_decay}
\end{equation}
Each CNP triggers a multiplicative decrease of the current rate:
\begin{equation}
R_C \leftarrow R_C\left(1 - \frac{\alpha}{2}\right)
\label{eq:rate_decrease}
\end{equation}
and, optionally, a clamp of the target rate to its pre-CNP value:
\begin{equation}
R_T \leftarrow R_C \text{ (pre-CNP)}.
\label{eq:clamp_tgt}
\end{equation}
The assignment~(\ref{eq:clamp_tgt}) is configurable: when left enabled through the ClampTargetRate option (the default in the DCQCN specification), the target rate is clamped to the pre-CNP current rate at each CNP, so that the algorithm forgets the previous high-rate target; when disabled, $R_T$ is left unchanged on CNPs, preserving a memory of the past target.
In the absence of CNPs, the recovery dynamics is driven by two independent triggers: a timer that fires periodically and a byte counter that fires once a programmable number of bytes has been transmitted since the last CNP. Each firing produces a \emph{recovery event} that increments a stage counter $F$ and applies one rate update; $F$ is reset to zero on every CNP. The functional form of the update depends on the value of $F$ at that event:
\begin{itemize}
\item in the \emph{Fast Recovery} phase ($F$ small) only the current rate is moved halfway toward the target, $R_C \leftarrow (R_C + R_T)/2$;
\item in the \emph{Additive Increase} phase ($F$ larger) the target rate is first incremented, $R_T \leftarrow R_T + R_{AI}$, then the same halving step is applied;
\item in the \emph{Hyper-Additive Increase} phase ($F$ larger still) the increment $R_{AI}$ is replaced by the larger $R_{HAI}$.
\end{itemize}

The qualitative behaviour produced by~(\ref{eq:alpha_up})--(\ref{eq:rate_decrease}) and by the three recovery phases is a per-flow saw-tooth around the bottleneck fair-share rate, with the depth of the saw-tooth controlled by $\alpha$ and the slope of the recovery controlled by the timer period, by the stage-counter transition thresholds, and by the magnitude of the increments $R_{AI}$ and $R_{HAI}$.

\section{Why PFC Alone Is Not Enough}
\label{sec:pfc}
The AdvCam DAQ is, by construction, an incast funnel: hundreds of front-end boards push data at $\sim$10\,Gb/s each into a smaller number of switch uplinks that converge onto a few high-speed server NICs. Unlike typical accelerator-based HEP experiments, where data generation is paced by the machine clock, the LST trigger is driven by atmospheric Cherenkov showers and therefore fires at random times. The data traffic through the switch is consequently unpredictable, and bursts in which many boards happen to transmit simultaneously cannot be excluded a priori. This makes the AdvCam fabric intrinsically prone to congestion. Without any congestion-control mechanism, the slightest mismatch between offered load and uplink capacity fills the switch buffer, packets are dropped, and the RoCEv2 reliability layer reacts with Go-Back-N retransmissions; as a consequence, throughput collapses.

PFC is the classical lossless mechanism. When an ingress queue on a switch port approaches its threshold, the switch emits an IEEE 802.1Qbb PAUSE frame upstream, halting the immediate sender (or senders) on the Layer-2 traffic class carrying the incoming traffic. PFC successfully prevents losses, but it operates at the granularity of a physical link and a traffic class. Two consequences follow. First, an innocent flow that happens to share an upstream link with a guilty one is paused as well (head-of-line blocking). Second, because PFC equalizes the bandwidth share per link rather than per flow, in any topology in which different switch ports aggregate different numbers of contending flows, those flows that sit on the busier port systematically receive less bandwidth than the others --- the classical \emph{parking-lot problem}.

\begin{figure}[t]
\centering
\includegraphics[width=0.85\columnwidth]{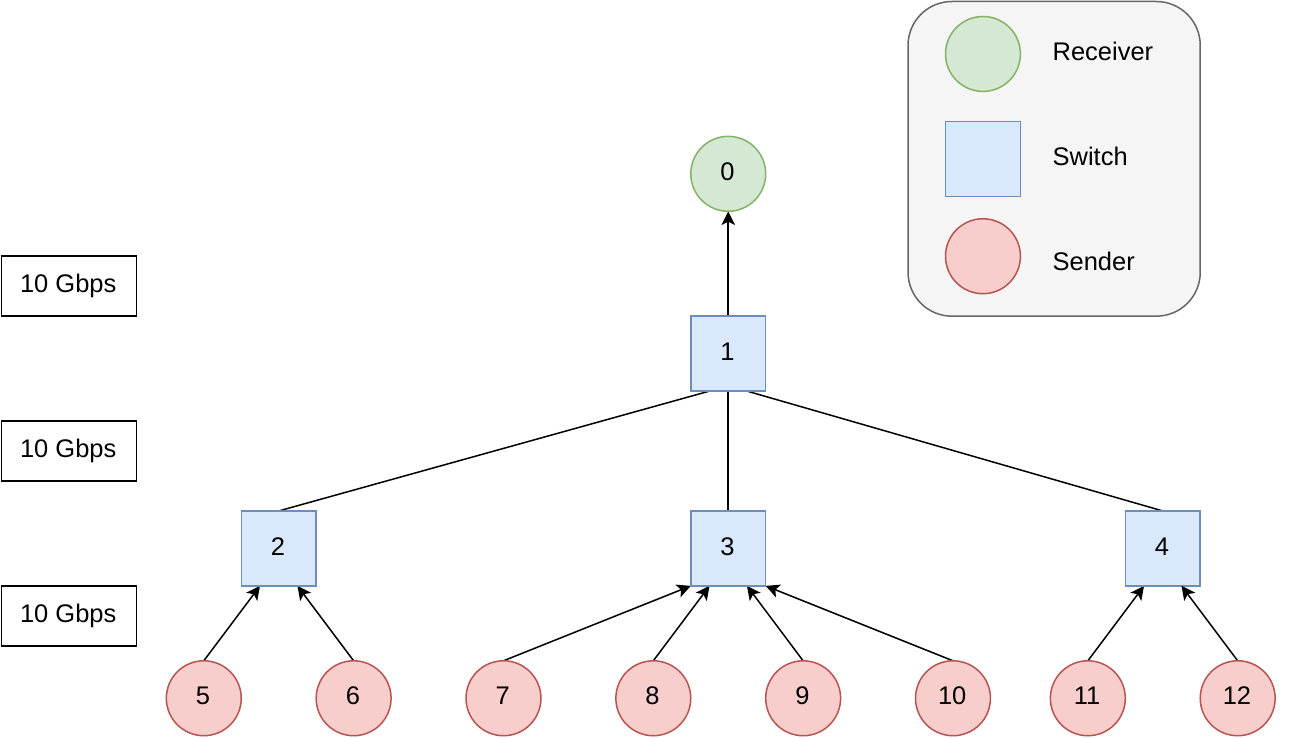}
\caption{Simulated AdvCam-like tree topology: eight senders (5--12) feed three leaf switches (2, 3, 4); leaf 3 aggregates four senders, leaves 2 and 4 only two each. The leaves converge on a single root switch (1) and one receiver (0).}
\label{fig:dc_topology}
\end{figure}

A simple ns-3 simulation on the AdvCam-like tree of Fig.~\ref{fig:dc_topology} illustrates the effect. Eight senders (nodes 5--12) target a single receiver (node 0) through two tiers of switches: a root switch (node 1) connected to the receiver, and three leaf switches (nodes 2, 3, 4) below it. The leaf switches are loaded asymmetrically: leaf 2 and leaf 4 are connected to two senders each (node 5--6 and node 11--12, respectively), while leaf 3 is connected to four senders (nodes 7--10). This is a deliberately simplified version of the structure that the AdvCam DAQ is expected to adopt, with front-end boards grouped on a per-cluster basis behind aggregation switches whose uplinks converge toward the event-building servers. In the final readout only the front-end-to-leaf links will run at 10\,Gb/s, while the leaf-to-root and root-to-server links will be upgraded to 100, 200, or 400\,Gb/s as the technology allows; in the simulation, however, all links are kept at 10\,Gb/s so that congestion can be exercised on a small, easy-to-interpret topology.

\begin{figure}[t]
\centering
\includegraphics[width=0.85\columnwidth]{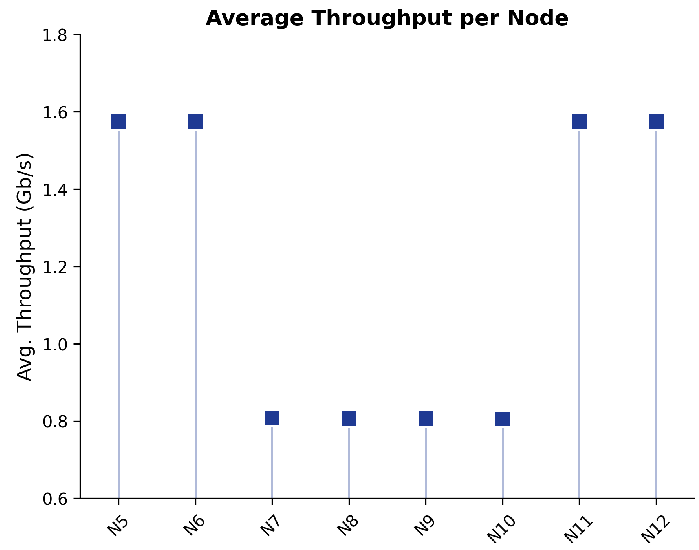}
\caption{Per-sender throughput with PFC only, ECN disabled. Senders behind the two-sender leaves (5--6, 11--12) obtain $\sim$1.58\,Gb/s, while those behind the four-sender leaf (7--10) obtain only $\sim$0.81\,Gb/s --- a textbook parking-lot effect.}
\label{fig:pfc_unfair}
\end{figure}

With ECN disabled in the switches and only PFC active, the per-sender throughput is distributed as in Fig.~\ref{fig:pfc_unfair}. PFC equalizes the bandwidth share at the root switch across the three incoming uplinks (each gets one third of the 10\,Gb/s receiver link, i.e.\ $\sim$3.3\,Gb/s). At leaves 2 and 4, two senders divide that share and each obtains $\sim$1.58\,Gb/s; at leaf 3, four senders divide the same share and each obtains $\sim$0.81\,Gb/s --- about half. The result is a textbook parking lot: the per-flow rate depends on which leaf a sender happens to sit behind, not on the offered traffic of that sender. For a DAQ in which all front-end boards carry physically meaningful detector data, this is plainly unacceptable.

DCQCN solves this by sliding the reaction from per-link to per-flow: it is the offending sender, identified by its CNP, that slows down, not whichever sender happens to be upstream of a congested link.

\section{Protocol-Level Validation in ns-3}
\label{sec:ns3}
Before committing the FPGA development effort, the DCQCN dynamics were validated at the protocol level in ns-3, using the open-source \texttt{ns3-roce} module~\cite{ns3roce}.

\subsection{Simulation Setup}
The topology, the link rates, and the placement of senders are the same as in Sec.~\ref{sec:pfc} (Fig.~\ref{fig:dc_topology}); only the switch configuration is changed, with DCQCN now enabled in addition to PFC.

The ECN marking profile follows a standard Random Early Detection (RED) curve, with $K_{min}$, $K_{max}$ and $p_{max}$ adapted to the line rate of each interface, and the DCQCN parameters follow the recommended values of~\cite{zhu2015dcqcn}.

\subsection{Results}
\begin{figure}[t]
\centering
\includegraphics[width=0.85\columnwidth]{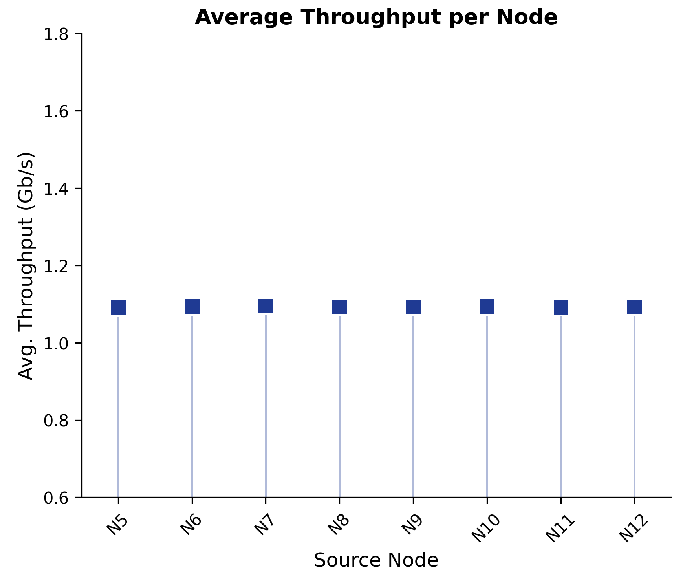}
\caption{Per-sender throughput with PFC and DCQCN enabled: all eight flows converge to the same $\sim$1.1\,Gb/s share, regardless of the leaf switch they sit behind.}
\label{fig:dcqcn_fair_ns3}
\end{figure}

With DCQCN enabled, the per-sender throughput of the eight contending flows collapses on a single fair-share value of $\sim$1.1\,Gb/s, with negligible spread (Fig.~\ref{fig:dcqcn_fair_ns3}). The parking-lot asymmetry of the PFC-only baseline (Fig.~\ref{fig:pfc_unfair}) is removed: the four senders behind leaf 3 now receive the same rate as the two senders behind leaves 2 and 4. The aggregate of the eight flows tracks the receiver's 10\,Gb/s link with no losses recorded.

\begin{figure}[t]
\centering
\includegraphics[width=0.95\columnwidth]{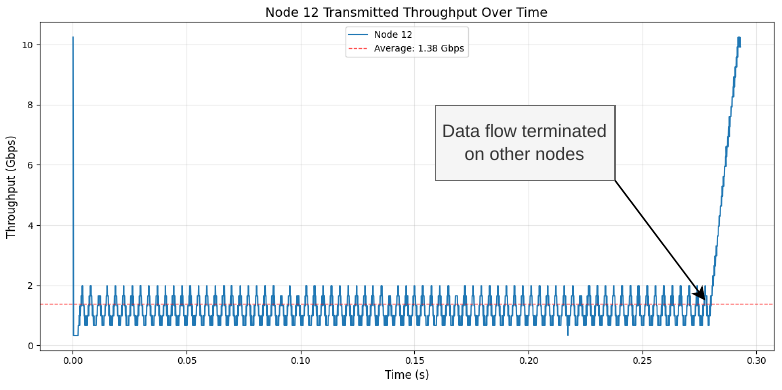}
\caption{Instantaneous throughput of one sender (node~12) in the ns-3 DCQCN run: oscillations around the fair-share rate during contention, sharp climb to the line rate as the other flows complete.}
\label{fig:dcqcn_time_ns3}
\end{figure}

The single-flow time profile of Fig.~\ref{fig:dcqcn_time_ns3} shows the canonical DCQCN behaviour for one of the senders: small, fast oscillations around the fair-share rate during the contention phase, followed by a sharp climb toward the line rate as the other flows complete (visible after $t \approx 0.28$\,s).

This experiment served two purposes. First, it confirmed that the DCQCN parameter family adopted from~\cite{zhu2015dcqcn} produces the expected dynamics on the AdvCam-like topology and rate regime, before any FPGA work was committed. Second, even though the FPGA parameters are run-time programmable through the register interface mentioned in Sec.~\ref{sec:fpga} (no synthesis cycle is needed to change them), the simulator allowed a structured, instrumented exploration of the parameter space in scenarios that would be tedious to set up on hardware --- in particular many-sender configurations with controllable per-flow start times --- so that the hardware bring-up could start from an informed operating point rather than from random trials.

\section{FPGA Implementation}
\label{sec:fpga}
\begin{figure}[t]
\centering
\includegraphics[width=0.98\columnwidth]{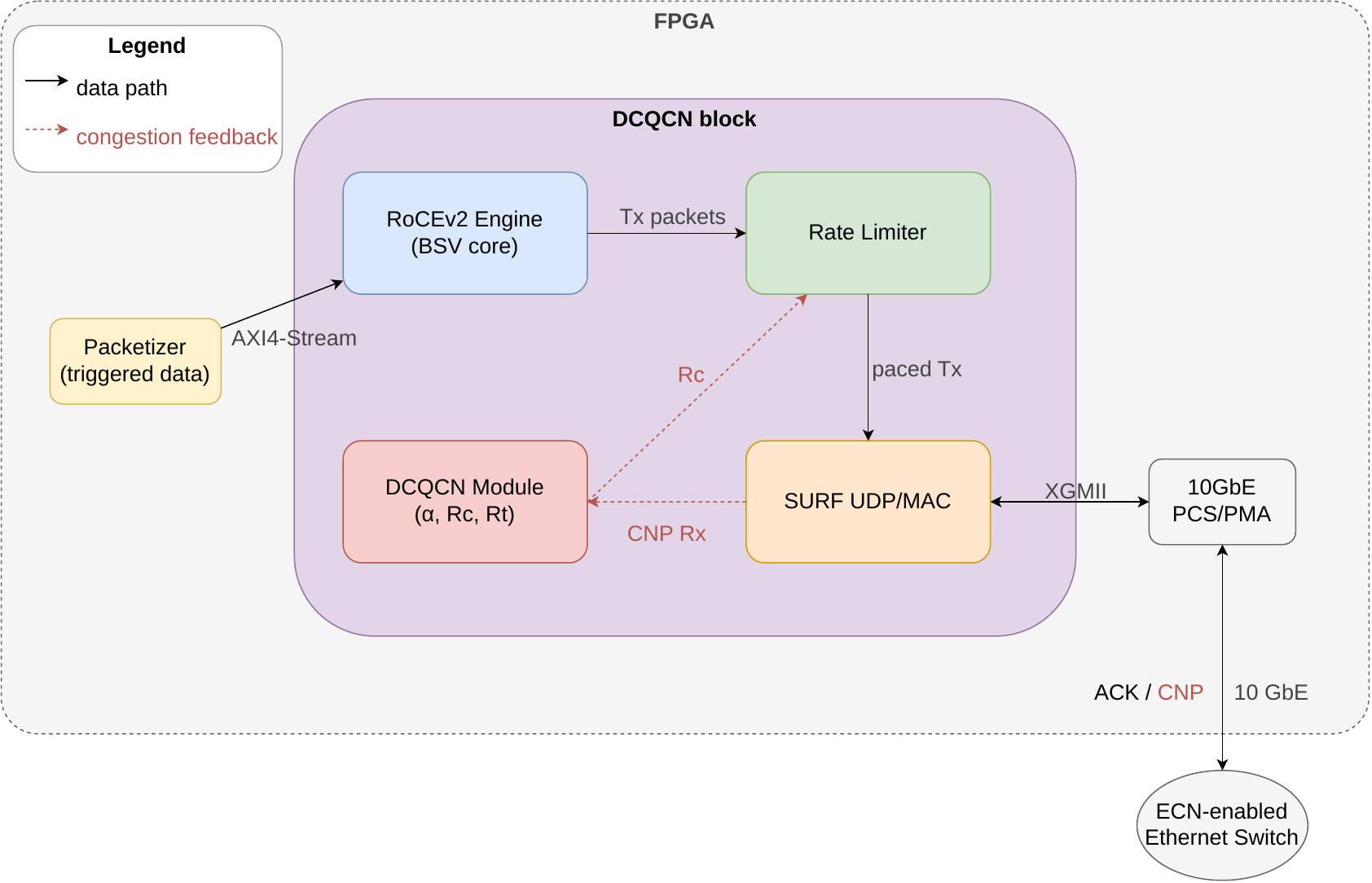}
\caption{FPGA network stack with the DCQCN rate limiter inserted between the RoCEv2 engine and the SURF UDP/IP+MAC.}
\label{fig:fpga_block}
\end{figure}

The DCQCN reaction is implemented as a VHDL module inserted on the AXI4-Stream datapath between the RoCEv2 engine and the customized SURF UDP/IP+MAC stack of~\cite{marini2025fpga}. The full network stack in the FPGA, including the DCQCN block, is shown in Fig.~\ref{fig:fpga_block}. The choice of VHDL, rather than the Bluespec SystemVerilog of the original RoCEv2 core, was driven by the better integration with the surrounding SURF library, which is itself VHDL-based.

The module exposes an AXI4-Stream bus for the network traffic and a CNP reception port that is asserted whenever the inbound MAC path identifies an incoming CNP packet. A rising-edge synchronizer feeds the asserted CNP into three concurrent sub-processes:
\begin{itemize}
\item the \emph{alpha update} process, which implements~(\ref{eq:alpha_up}) and~(\ref{eq:alpha_decay}) at a configurable update interval;
\item the \emph{rate decrease} process, which implements~(\ref{eq:rate_decrease}) on each CNP rising edge, with a programmable cooldown to prevent reacting twice to the same congestion epoch;
\item the \emph{rate increase} process, which implements the three recovery phases (Fast Recovery, Additive Increase, Hyper-Additive Increase) using a programmable timer, a programmable byte counter, and the stage counter $F$ introduced in Sec.~\ref{sec:dcqcn}.
\end{itemize}

The rate limiter itself uses $R_C$ as a credit generator: a fractional credit accumulator increments by $R_C / f_{\text{clk}}$ per cycle (in bytes) and is decremented by the number of payload bytes that move on the AXI4-Stream master in the same cycle. When the accumulator is below the size of the next AXI4-Stream beat, the master is back-pressured. Because the SURF MAC is the sole downstream consumer, the egress link rate is governed exclusively by $R_C$, with no hidden bottlenecks.

All DCQCN parameters --- $g$, $R_{AI}$, $R_{HAI}$, the alpha-update interval, the rate-decrease interval, the rate-increase timer period, the byte-counter threshold, the stage-counter transition threshold, and the ClampTargetRate flag --- are mapped onto registers that are exposed at run time through the native register-access interface provided by the SURF library, on top of which the module is built. The same interface exports the live $R_C$, $R_T$, and $\alpha$ registers for monitoring, a feature exploited in Sec.~\ref{sec:hw} to overlay the rate-limiter state on the measured throughput.

The resource utilization of the DCQCN module, after synthesis on a Xilinx/AMD XCKU040 FPGA, is reported in Table~\ref{tab:dcqcn_resources}. The footprint is negligible on top of the modified RoCEv2 core of~\cite{marini2025fpga}: less than 0.5\% of the available LUTs and FFs, well below 1\% of the DSP slices.

\begin{table}[t]
\caption{DCQCN module resource utilization on XCKU040 after synthesis.}
\label{tab:dcqcn_resources}
\centering
\begin{tabular}{lrr}
\toprule
Resource & Used & Percentage \\
\midrule
LUT  & 1337 & 0.5\% \\
FF   & 2557 & 0.5\% \\
BRAM & 18   & 3.0\% \\
DSP  & 5    & 0.3\% \\
\bottomrule
\end{tabular}
\end{table}

\section{Cocotb Firmware Simulation}
\label{sec:sim}
\begin{figure}[t]
\centering
\includegraphics[width=0.75\columnwidth]{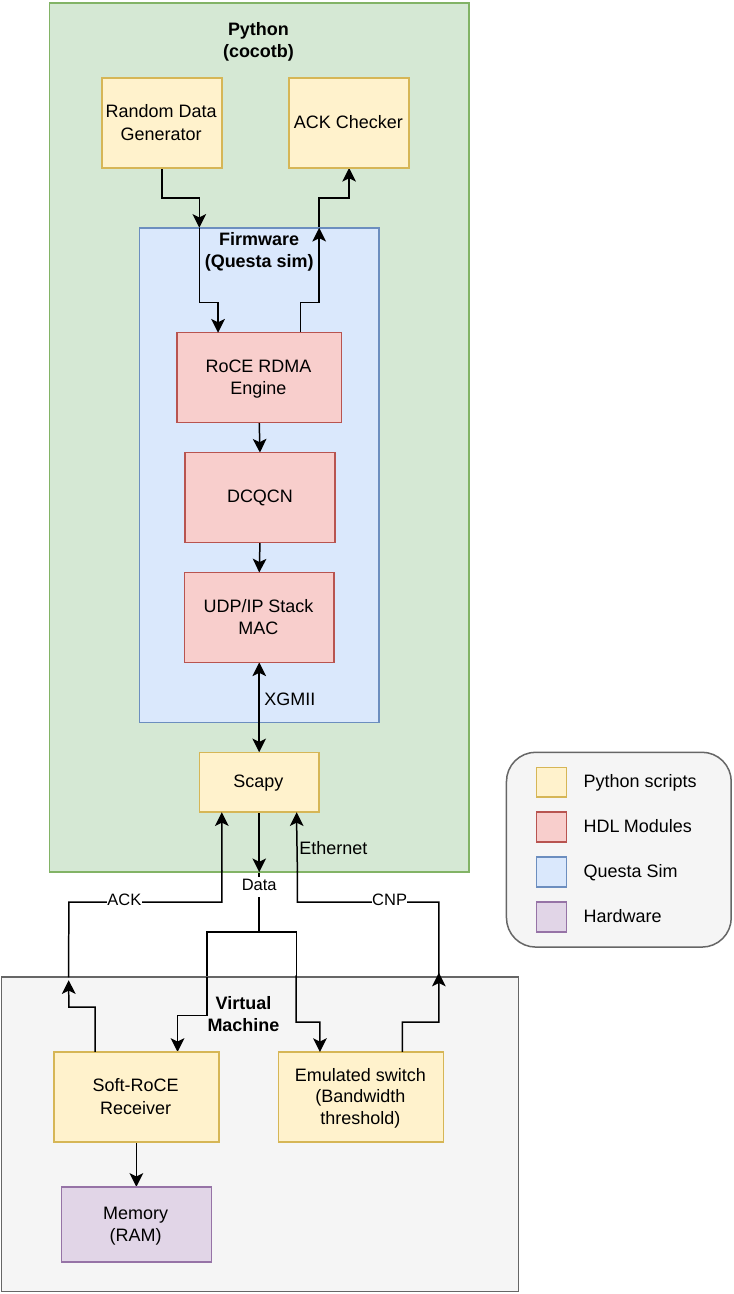}
\caption{DCQCN firmware simulation architecture: cocotb drives the HDL stack in Questa; Scapy bridges XGMII to a Soft-RoCE VM; an emulated switch monitors throughput and forges CNPs back when a threshold is exceeded.}
\label{fig:dcqcn_sim_block}
\end{figure}

The DCQCN firmware was verified end-to-end against a real RoCEv2 receiver using a cocotb~\cite{cocotb} testbench. cocotb is a Python-based co-simulation framework that drives an HDL simulator --- Questa in our case --- through Python coroutines, so the FPGA can be exercised by arbitrary high-level code rather than by hand-written HDL stimuli. The architecture of the simulation is shown in Fig.~\ref{fig:dcqcn_sim_block}. The cocotb testbench instantiates the full RoCEv2+DCQCN+SURF stack inside Questa and drives its input AXI4-Stream with synthetic payload. The outbound XGMII frames are parsed with \textit{scapy}~\cite{rohith2018scapy} --- a Python library for packet manipulation --- and forwarded over the network to a virtual machine running Soft-RoCE~\cite{softroce}, the Linux kernel software RoCEv2 implementation, which treats the simulated FPGA as a legitimate remote RDMA peer: it completes the RoCEv2 handshake, accepts RDMA WRITE payload into a pre-registered memory region, and replies with ACKs that travel back into the simulation along the same path. Further details on the original verification strategy are given in~\cite{marini2025fpga}. For DCQCN, two ingredients are added on the receive side: an emulated switch, written in Python, that interposes between the simulated FPGA and the Soft-RoCE receiver, and a runtime dashboard that monitors the simulation state through a local web server.

\subsection{Emulated Switch}
The emulated switch is a cocotb coroutine that records every payload byte observed on the XGMII transmit bus and computes the instantaneous throughput over a sliding window of $N_w$ clock cycles, $T_{\text{win}} = N_w \times T_{\text{clk}}$. With $N_w = 5000$ and $T_{\text{clk}} = 6.4$\,ns this corresponds to a measurement window of 32\,$\mu$s. Whenever the measured throughput exceeds a programmable threshold and a cooldown interval has elapsed since the last trigger, the emulated switch pushes a bandwidth-reduction trigger to a listener running inside the Soft-RoCE virtual machine. The listener, in turn, forges and injects a properly addressed CNP packet back into the simulation through the same network path used for ACKs.

This design has two attractive properties for verification. First, it provides a genuine end-to-end test of the DCQCN feedback loop: the Soft-RoCE virtual machine consumes ordinary network packets and is, by construction, unaware that they originate from a simulated FPGA rather than from a real RDMA-capable NIC. The full RoCEv2 exchange --- payload, ACKs, and CNP feedback --- is therefore exercised and any deviation from the protocol on the FPGA side is immediately exposed by the receiver's reaction. Second, the threshold can itself be scripted as a function of simulation time, allowing scenarios in which the available bandwidth at the bottleneck is varied during the run, mimicking the arrival and departure of contending flows in a real incast.

\subsection{Results}
\begin{figure}[t]
\centering
\includegraphics[width=\columnwidth]{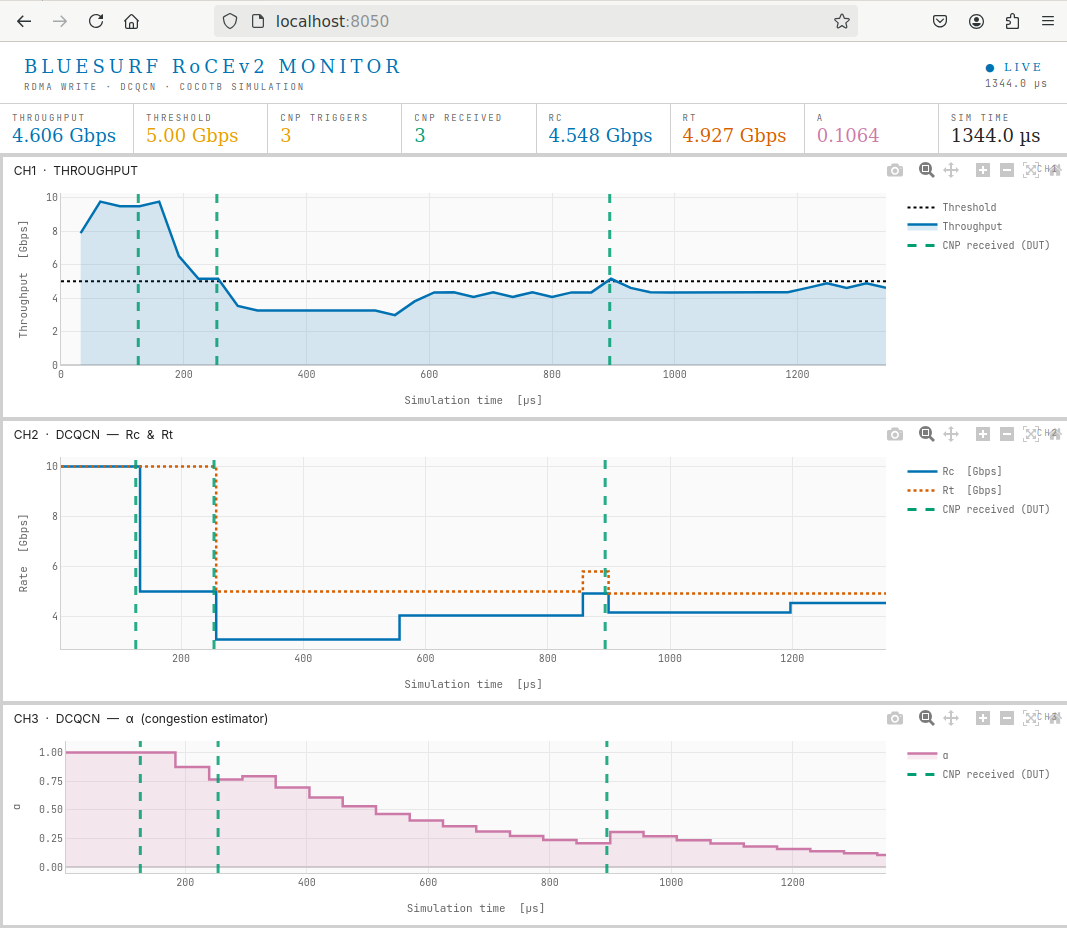}
\caption{DCQCN cocotb simulation dashboard. Top: measured throughput vs.\ emulated-switch threshold. Middle: $R_C$ and $R_T$. Bottom: congestion estimator $\alpha$. Dashed green vertical lines mark CNPs received by the FPGA.}
\label{fig:dcqcn_sim_dash}
\end{figure}

A representative run is shown in Fig.~\ref{fig:dcqcn_sim_dash}. The emulated switch is set to a 5\,Gb/s threshold; the FPGA starts unconstrained at the 10\,Gb/s line rate. Three CNPs arrive over the run, each catching $\alpha$ at a different value: the first, with $\alpha=1$, halves $R_C$ from 10 to 5\,Gb/s; the second, with $\alpha \approx 0.8$, produces a smaller step; the third, after $\alpha$ has decayed to $\approx 0.3$, barely perturbs $R_C$. Between CNPs, the rate-increase process climbs $R_C$ in Fast Recovery steps toward $R_T$ and the throughput tracks $R_C$ to within the measurement window.

In all simulated runs, the post-run inspection of the receiver memory and of the FPGA Completion Queue confirmed that no Work Request was lost or completed with an error, even when the threshold was lowered aggressively to provoke many CNPs in quick succession.

\section{Hardware Measurements}
\label{sec:hw}
\begin{figure}[t]
\centering
\includegraphics[width=0.95\columnwidth]{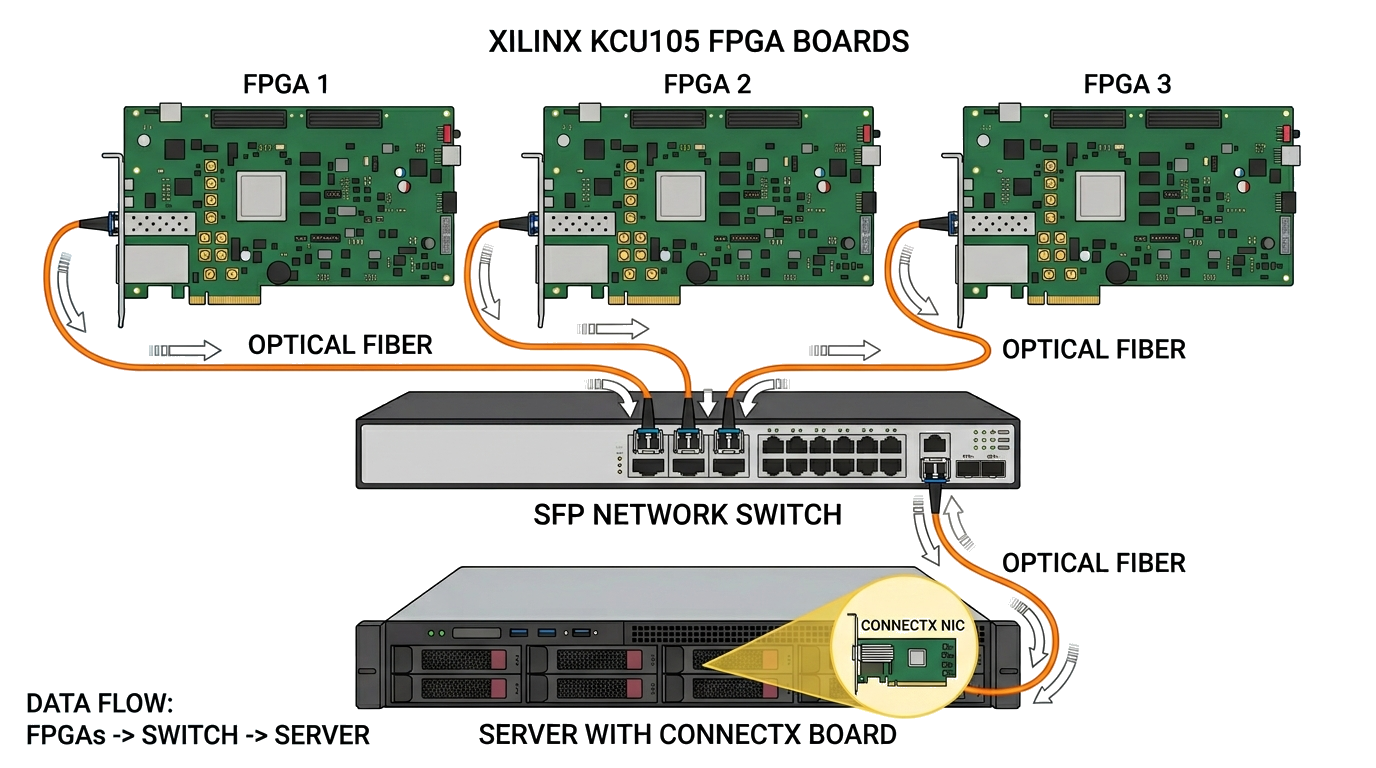}
\caption{Hardware test bed: three KCU105 RoCEv2+DCQCN senders, one ECN-enabled FS N5860-48SC switch, one ConnectX-5 receiver. All links at 10\,Gb/s.}
\label{fig:hw_setup}
\end{figure}

The DCQCN-enabled firmware was then exercised on real hardware in the incast configuration of Fig.~\ref{fig:hw_setup}. Three Xilinx/AMD KCU105 evaluation boards~\cite{kcu105}, each carrying the same modified RoCEv2+DCQCN+SURF firmware, act as RoCEv2 senders. Each board is connected at 10\,Gb/s to an FS N5860-48SC commercial Ethernet switch~\cite{fs5860}, configured with ECN marking on the egress queue toward the server uplink and with PFC enabled on the traffic class carrying the RDMA traffic. The server uplink is 10\,Gb/s, terminated by a Mellanox ConnectX-5 NIC~\cite{connectx5} acting as the single receiver. The classical incast condition is realized: $3 \times 10$\,Gb/s offered, $1 \times 10$\,Gb/s sink.

\subsection{DCQCN Parameter Set}
Unless otherwise stated, the three senders are programmed with the same DCQCN parameters, summarized in Table~\ref{tab:hw_params}. The values were selected starting from the ns-3 results of Sec.~\ref{sec:ns3} and then refined empirically on hardware. They are written to each board through the run-time register interface of the SURF network stack mentioned in Sec.~\ref{sec:fpga}, before the test begins.

\begin{table}[t]
\caption{DCQCN parameters used in the three-sender incast hardware test.}
\label{tab:hw_params}
\centering
\begin{tabular}{lr}
\toprule
Parameter & Value \\
\midrule
Gain $g$                   & $1/2^8 \approx 3.9 \times 10^{-3}$ \\
Additive increment $R_{AI}$   & 6\,MB/s \\
Hyper-additive incr.\ $R_{HAI}$ & 12\,MB/s \\
Rate-decrease interval     & 3\,$\mu$s \\
$\alpha$-update interval   & 40\,$\mu$s \\
Rate-increase interval     & 2\,ms \\
Stage-counter threshold $F$ & 5 \\
ClampTargetRate            & on \\
MTU                        & 4096\,B \\
\bottomrule
\end{tabular}
\end{table}

\subsection{Without DCQCN}
\begin{figure}[t]
\centering
\includegraphics[width=0.95\columnwidth]{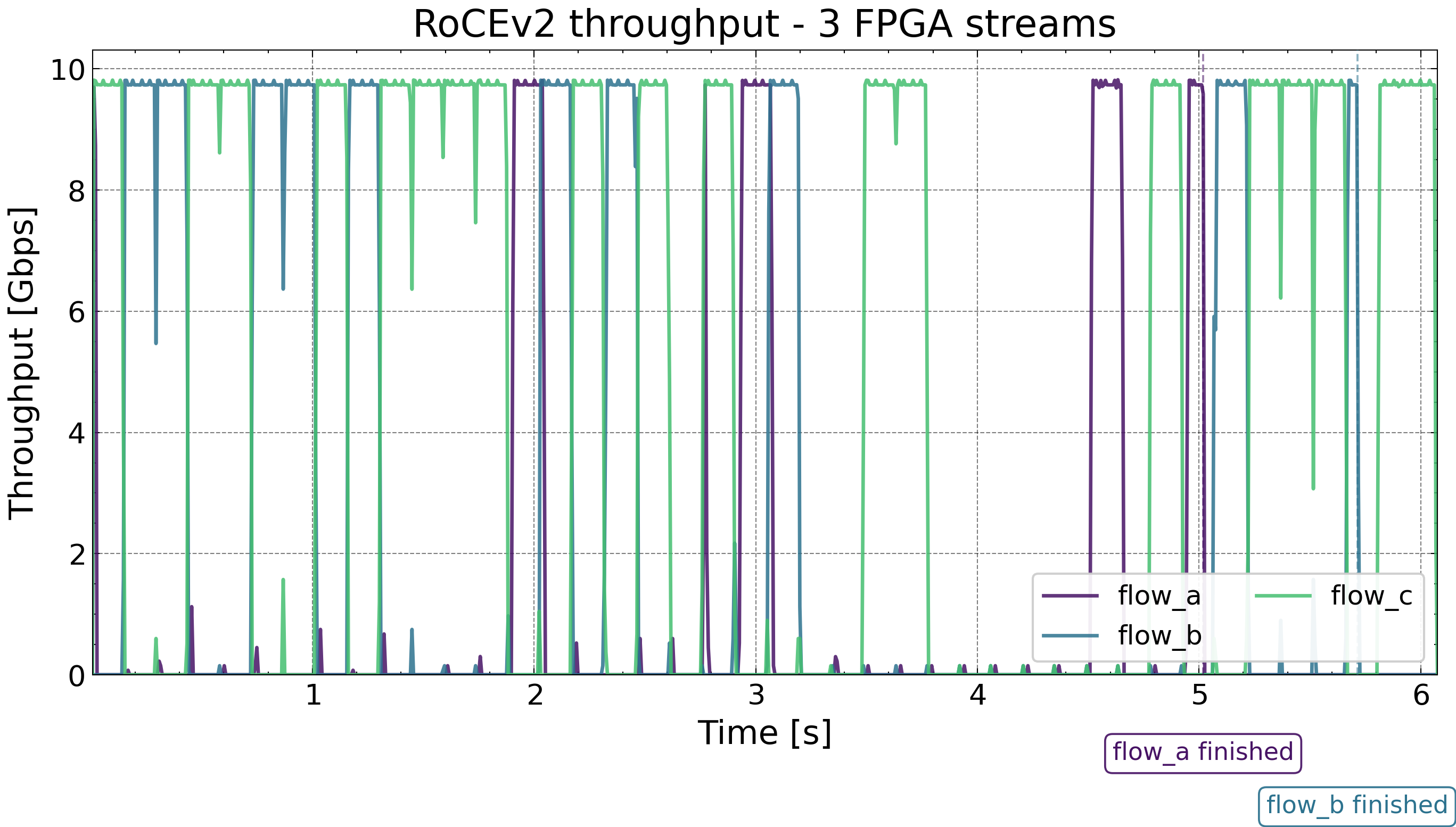}
\caption{Three-sender incast, DCQCN disabled: chaotic on/off pattern, unfair allocation.}
\label{fig:no_dcqcn}
\end{figure}

As a baseline, the rate-limiter reaction was disabled (the DCQCN module passes data through unconditionally and ignores CNPs) and the same three-sender incast was repeated. The result is shown in Fig.~\ref{fig:no_dcqcn}. The pattern is chaotic: each sender alternates between long bursts at the full 10\,Gb/s line rate and intervals of essentially zero throughput, during which it is paused by PFC or stalled by RoCEv2 retransmissions. The instantaneous aggregate frequently approaches the line rate but the per-flow share is patently unfair and bursty.

\subsection{With DCQCN}
\begin{figure}[t]
\centering
\includegraphics[width=0.95\columnwidth]{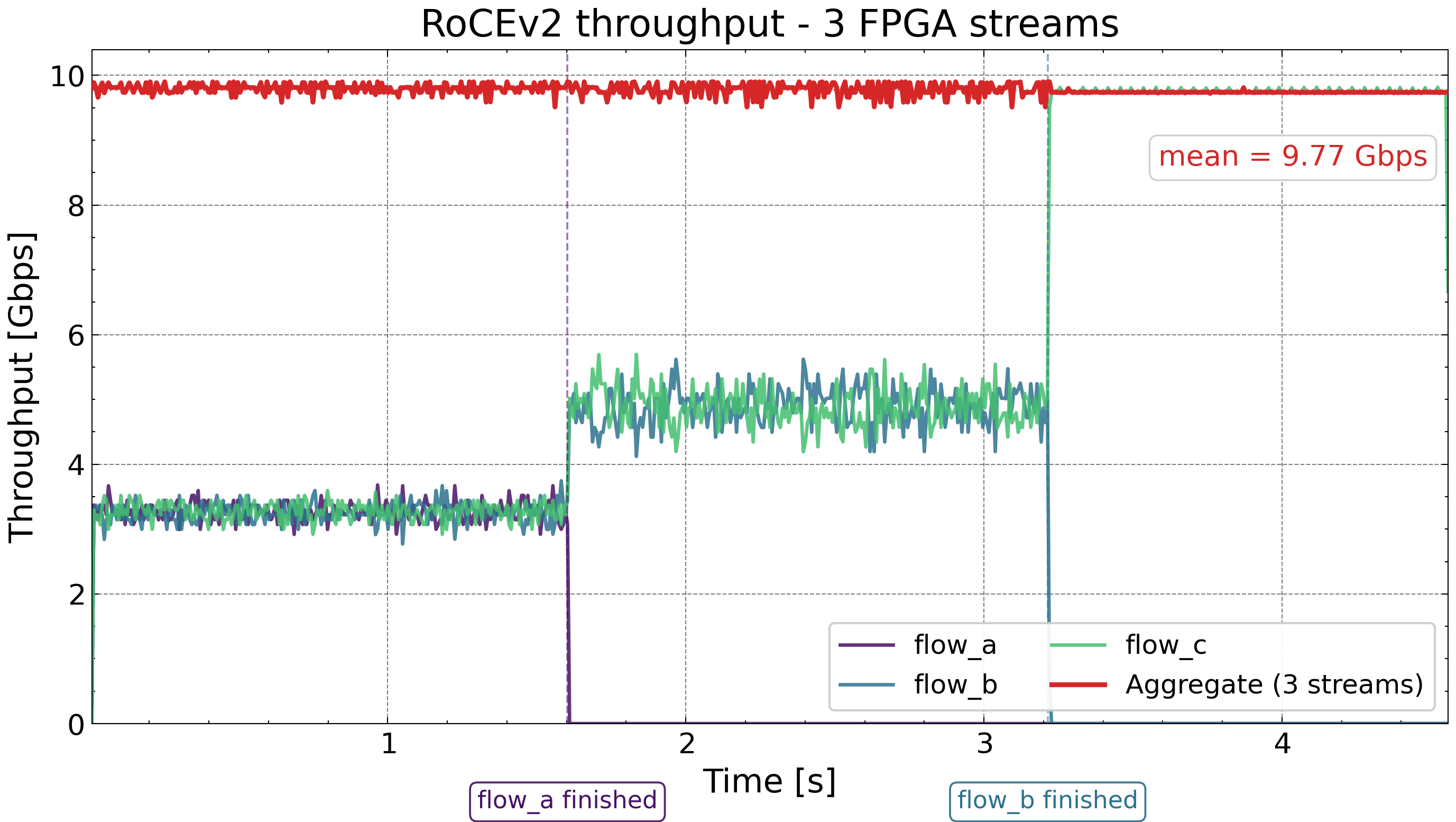}
\caption{Three-sender incast, DCQCN enabled (parameters from Table~\ref{tab:hw_params}): fair $\sim$3.3\,Gb/s share, reactive redistribution as senders finish, 9.77\,Gb/s aggregate.}
\label{fig:dcqcn_hw_fair}
\end{figure}

With DCQCN enabled and identical parameters on all three senders, the same test produces the trace of Fig.~\ref{fig:dcqcn_hw_fair}. The three flows converge promptly to a smooth, near-equal share of approximately 3.3\,Gb/s each. As the first sender (flow\_a) terminates, the remaining two redistribute the freed bandwidth and stabilize around 4.9\,Gb/s each; when the second sender terminates, the surviving flow climbs back to the full $\sim$9.7\,Gb/s line rate. The aggregate throughput averages 9.77\,Gb/s for the entire run, no packet losses are recorded, and the redistribution between the three plateaus is reactive on a sub-second time scale.

\subsection{Asymmetric Parameters: Visualizing the Saw-Tooth}
The symmetric configuration of Fig.~\ref{fig:dcqcn_hw_fair} produces a textbook fair share but, precisely because the three flows oscillate in phase opposition around the same mean, the individual rate dynamics that each rate limiter executes is largely washed out in the instantaneous throughput trace. To make the state machine visible directly in the hardware data, a deliberately asymmetric run was performed: one sender (flow\_a) was configured with a rate-increase timer of $\sim$200\,ms, two orders of magnitude longer than the 2\,ms used by the other two, and with the ClampTargetRate option turned off so that $R_T$ is preserved across CNPs. The other two senders kept the values of Table~\ref{tab:hw_params}.

\begin{figure}[t]
\centering
\includegraphics[width=0.95\columnwidth]{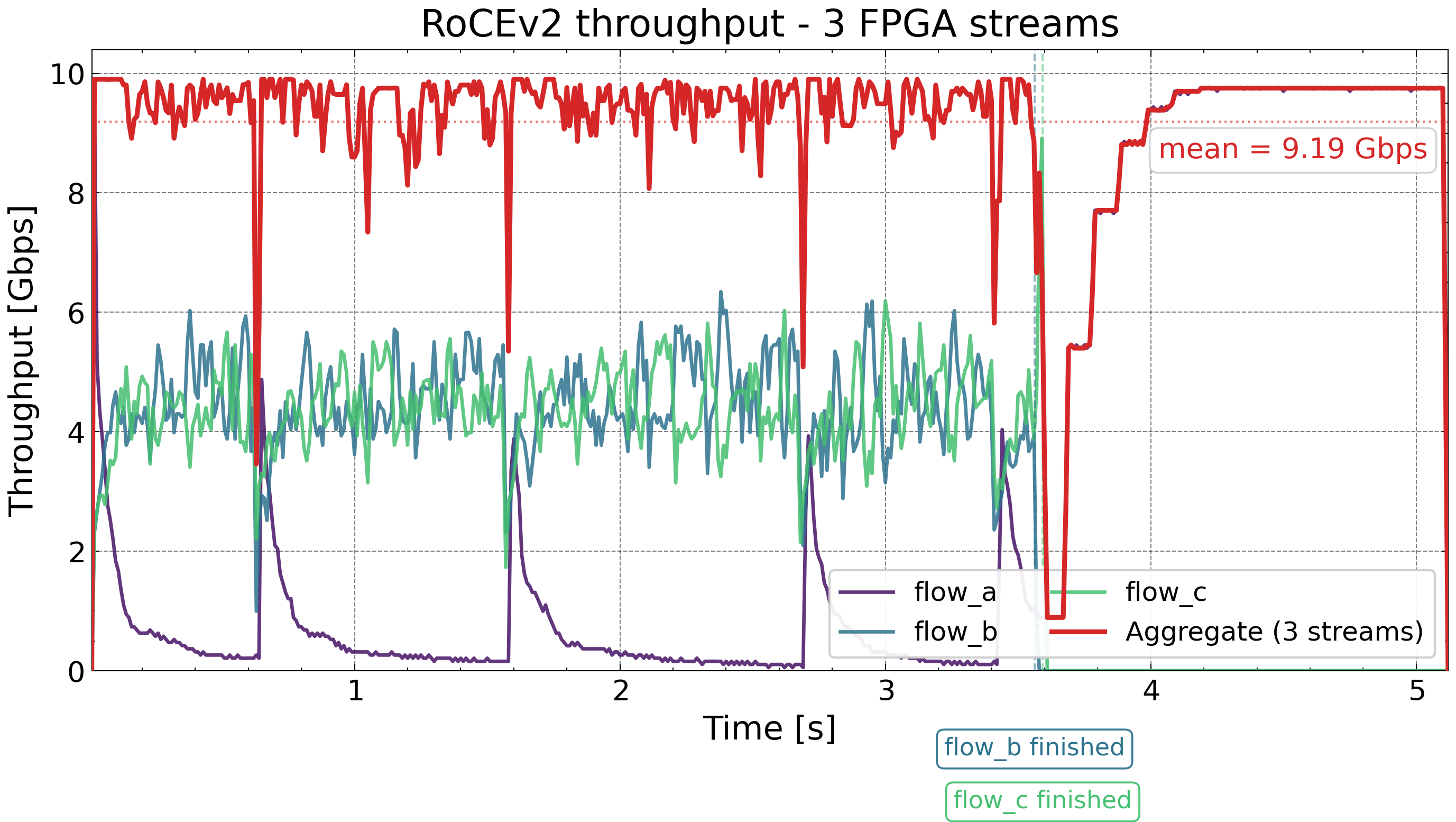}
\caption{Asymmetric DCQCN parameters: flow\_a uses a 200\,ms rate-increase timer and ClampTargetRate off, exposing on its trace a sharp climb followed by exponential decay characteristic of that configuration.}
\label{fig:dcqcn_hw_saw}
\end{figure}

The result is shown in Fig.~\ref{fig:dcqcn_hw_saw}. The slowed-down sender becomes the only one in which the recovery dynamics is slow enough to resolve in the throughput measurement window. Its trace shows a characteristic shape that is the combined signature of ClampTargetRate=off and a very long rate-increase timer: as soon as a recovery event finally fires, $R_C$ jumps up to $(R_C + R_T)/2$ with $R_T$ still holding the memory of a previous high value, producing a sharp climb toward the bottleneck rate. The high $R_C$ immediately re-induces congestion, a burst of CNPs arrives, and $R_C$ decays back toward zero in approximately exponential fashion over many CNPs, until the next recovery event triggers another upward jump. The two unmodified senders absorb the freed bandwidth in the meanwhile.

\subsection{Rate Limiter vs.\ Throughput}
\begin{figure}[t]
\centering
\includegraphics[width=0.95\columnwidth]{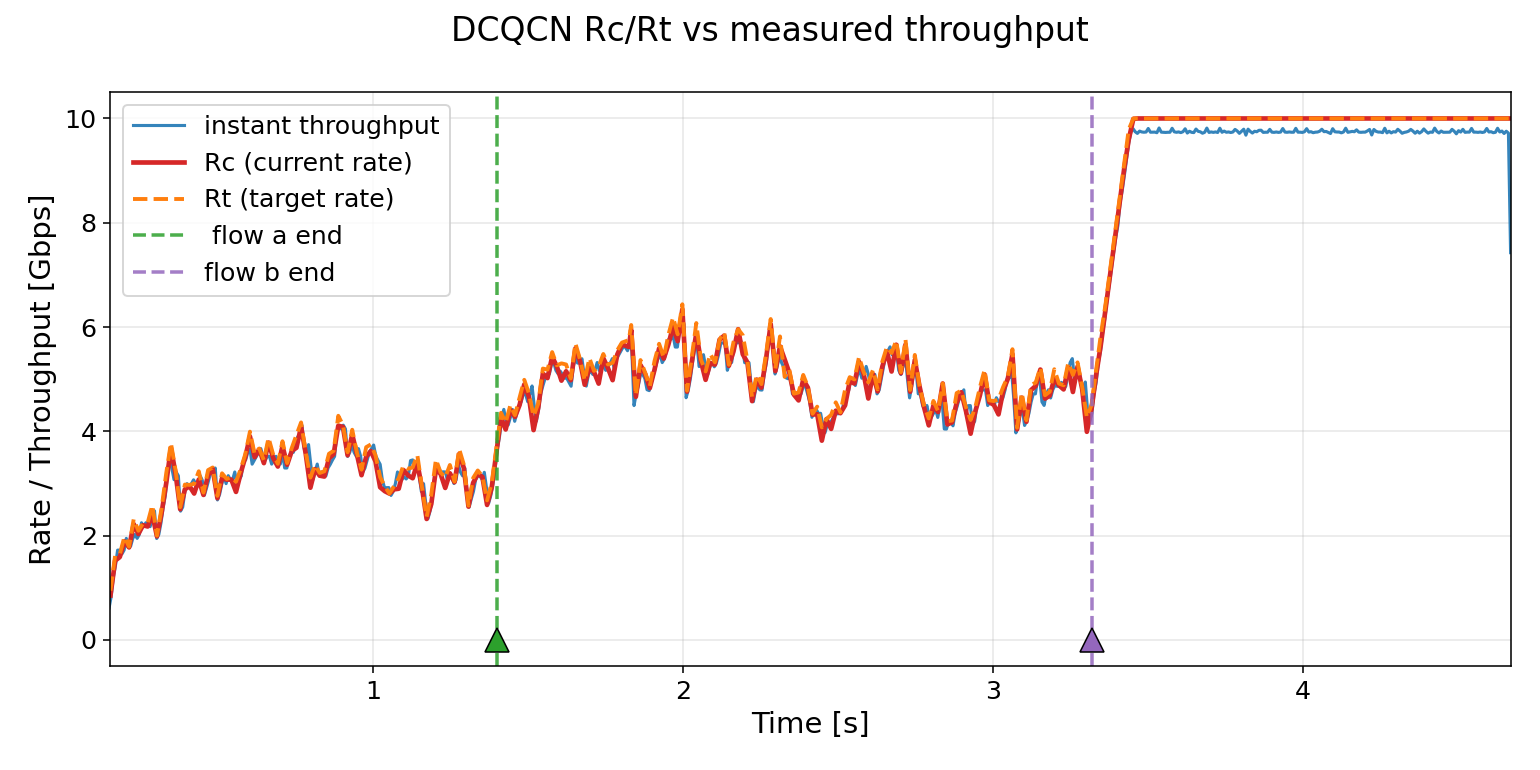}
\caption{Measured RoCEv2 throughput (blue) overlaid with DCQCN $R_C$ (red) and $R_T$ (dashed orange) for one sender. The three curves overlap throughout the run.}
\label{fig:overlay}
\end{figure}

A more quantitative cross-check exploits the monitoring path exposed by the DCQCN module through the SURF register interface mentioned in Sec.~\ref{sec:fpga}. The internal $R_C$ and $R_T$ registers were sampled at $\sim$1\,kHz during one of the three-sender runs and superimposed on the actual measured RoCEv2 throughput of the same sender (Fig.~\ref{fig:overlay}). The three curves --- instantaneous measured throughput (blue), $R_C$ (red), and $R_T$ (dashed orange) --- overlap throughout the run, including the two transitions where senders drop out. This confirms that the FPGA's actual link rate tracks the rate-limiter state cycle by cycle, with no hidden bottleneck in the SURF MAC, the PCS/PMA, or the optical link, and that the DCQCN block is the sole governor of the egress throughput.

\section{Conclusions and Outlook}
\label{sec:conclusion}
A complete DCQCN congestion-control implementation has been added to the FPGA RoCEv2 stack of~\cite{marini2025fpga}, addressing the open question of fair, lossless behaviour in an incast configuration representative of the AdvCam DAQ. The implementation was validated at three independent levels: protocol-level in ns-3, firmware-level in a cocotb simulation against a Soft-RoCE receiver and a Python emulated switch, and hardware-level on a three-sender incast test bed built around a commercial DCQCN-enabled switch and a ConnectX-5 server.

In hardware, with three FPGA RoCEv2 senders converging onto a single 10\,GbE receiver, the modified firmware reproduces the behaviour expected from a commercial RDMA NIC: a smooth and near-equal bandwidth allocation between contending flows, reactive redistribution as senders join or leave, no packet losses, and DCQCN recovery dynamics that emerge clearly when the parameters are skewed to expose them. The rate-limiter internal state was shown to track the measured throughput cycle by cycle, confirming that the DCQCN block is the sole governor of the egress link.

On the resource side, the DCQCN module adds less than 0.5\% of the LUTs and FFs of an XCKU040, an essentially free addition on top of the customized RoCEv2 core. The module exposes all algorithm parameters through an AXI-Lite interface, making it possible to re-tune the response at run time, a feature that has proven useful both for debugging and for adapting to different fabric configurations.

The next step is the integration of the DCQCN-aware firmware into the full AdvCam readout demonstrator, in which the FADC front-end of~\cite{marini2025fpga} is connected through an ECN-capable commercial switch to the event-building servers. The DCQCN module has been merged upstream into the SLAC SURF library, joining the iCRC contributions of~\cite{marini2025fpga}.

\bibliographystyle{IEEEtran}
\bibliography{refs}

\end{document}